# A Linear Branch Flow Model for Radial Distribution Networks and its Application to Reactive Power Optimization and Network Reconfiguration

Tianshu Yang, *Student Member, IEEE*, Ye Guo, *Senior Member*, *IEEE*, Lirong Deng, *Student Member, IEEE*, Hongbin Sun, *Fellow*, *IEEE*, and Wenchuan Wu, *Senior Member*, *IEEE*

***Abstract*—This paper presents a cold-start linear branch flow model named modified DistFlow. In modified DistFlow, the active and reactive power are replaced by their ratios to voltage magnitude as state variables, so that errors introduced by conventional branch flow linearization approaches due to their complete ignoring of network loss terms are reduced. Based on the path-branch incidence matrix, branch power flows and nodal voltage magnitudes can be written in succinct matrix forms. Subsequently, the proposed modified DistFlow model is applied to the problem of reactive power optimization and network reconfiguration, transforming it into a mixed-integer quadratic programming (MIQP). Simulations show that the proposed modified DistFlow has a better accuracy than existing cold-start linear branch flow models, and the resulting MIQP model for reactive power optimization and network reconfiguration is much more computationally efficient than existing benchmarks.**

## NOMENCLATURE

*Sets*

| | |
|---|---|
| $\Phi_B$ | Set of all buses |
| $\Phi_{Sub}$ | Set of all substations |
| $\Phi_G$ | Set of all DGs |
| $\Phi_C$ | Set of VAR compensators |
| $\Phi_L$ | Set of all branches |
| $\mathrm{N}_c(i)$ | Set of children buses of bus *i* |
| $\Psi_i$ | Set of branches on the path of bus *i* |
| $\Omega_{ij}$ | Set of the buses with branch *ij* in its path to root bus |
| $\mathrm{B}_l$ | Set of the branches in the fundamental loop *l* |
| $\Theta_i$ | Set of the branches in the *i*th overlapping loop set |

*Indices and Parameters*

| | |
|---|---|
| $l_{ij}$ | Branch between buses *i* and *j* |
| $R_{ij}, X_{ij}$ | Resistance and reactance of branch *ij* |
| $\tilde{P}_i^G, \tilde{Q}_i^G$ | Forecasted active and reactive power provided by DG at bus *i* |
| $P_i^D, Q_i^D$ | Active and reactive power demand at bus *i* |
| $N_{node}$ | Number of buses ( $N_{node} = \lvert\Phi_B\rvert$ ) |
| $N_{sub}$ | Number of substation buses |
| $N_G$ | Number of DGs |
| $N_i$ | Number of branches in *i*th overlapping loop set |
| $L_i$ | Number of links in *i*th overlapping loop set |
| $\bar{Q}_i^C, \underline{Q}_i^C$ | Maximum and minimum reactive power provided by SVC at bus *i* |
| $\bar{V}_i, \underline{V}_i$ | Upper and lower bound of voltage magnitude at bus *i* |
| $\bar{P}_{ij}, \bar{Q}_{ij}$ | Upper bound of active and reactive power flow on branch *ij* |
| $\bar{I}_{ij}$ | Upper bound of current of branch *ij* |
| $\bar{\delta}_{ij}$ | Upper bound of voltage angle difference of branch *ij* |
| $\underline{\eta}$ | Lower bound of power factor at feeder head |
| $M$ | A large positive number |

*Variables*

| | |
|---|---|
| $x_{ij}$ | Binary variable for branch status |
| $V_i$ | Voltage magnitude at bus *i* |
| $W_i$ | An auxiliary variable denoting $1/V_i$ |
| $\delta_{ij}$ | Voltage angle difference between buses *i* and *j* |
| $P_{ij}, Q_{ij}$ | Sending-end active and reactive power flow of branch *ij* |
| $P_i, Q_i$ | Total active and reactive power injection at bus *i* |
| $\hat{P}_{ij}, \hat{Q}_{ij}$ | Modified active and reactive power flow on branch *ij* |
| $\hat{P}_i, \hat{Q}_i$ | Modified active and reactive power injection at bus *i* |
| $\hat{P}_i^G, \hat{Q}_i^G$ | Modified active and reactive power injection provided by DG |
| $\hat{Q}_i^C$ | Modified reactive power injection provided by SVC |

*Vectors and Matrices*

| | |
|---|---|
| $\boldsymbol{T}$ | Path-branch incidence matrix for a radial network |
| $\boldsymbol{V}_S$ | Voltage magnitude vector of branch sending ends |
| $\boldsymbol{V}_R$ | Voltage magnitude vector of branch receiving ends with the same bus order as *T* matrix. |
| $\boldsymbol{W}_R$ | Vector of auxiliary variable $W_i$ |
| $\boldsymbol{V}_0$ | Column vector with all values equal to the voltage magnitude of PSP (the reference bus) |

This work was supported in part by the National Science Foundation of China under Grant 51977115. Corresponding author: H. Sun and Y. Guo.

T. Yang, Y. Guo, L. Deng, H. Sun and W. Wu are with the Smart grid and Renewable Energy Laboratory, Department of Tsinghua-Berkeley Shenzhen Institute (TBSI), Tsinghua University, Shenzhen 518055, Guangdong, China. (email: guo-ye@sz.tsinghua.edu.cn; shb@tsinghua.edu.cn).

H. Sun and W. Wu are also with the State Key Laboratory of Power Systems, Department of Electrical Engineering, Tsinghua University, Beijing 100084, China.

| | |
|---|---|
| $\hat{\boldsymbol{P}}_{Br}, \hat{\boldsymbol{Q}}_{Br}$ | Column vector of $\hat{P}_{ij}$ and $\hat{Q}_{ij}$ with the same branch order as $T$ matrix. |
| $\boldsymbol{P}_N, \boldsymbol{Q}_N$ | Diagonal matrix of active and reactive injected power ( $P_N = diag(P_i), Q_N = diag(Q_i)$ ) with the same bus order as $T$ matrix. |
| $\boldsymbol{R}_N, \boldsymbol{X}_N$ | Diagonal matrix of $R_{ij}$ and $X_{ij}$ with the same branch order as $T$ matrix. ( $R_N = diag(R_{ij}), X_N = diag(X_{ij})$ ) |

## I. Introduction

### A. Background and motivation

WITH the fast expansion of network scale and the deep penetration of renewable energy sources, increasingly complex and uncertain distribution systems need better control and optimization, such as network reconfiguration [1], reactive power (VAR) optimization [2], and economic dispatch [3]. To that end, the power flow model is an essential analytical tool. However, the standard alternating current power flow (ACPF) model is nonlinear and may make the aforementioned optimization problems non-convex.

Solving non-convex optimization is non-deterministic polynomial-time (NP) hard. There are two categories of methods to improve the computability for power flow optimizations: linearization on power flow equations and convex relaxation on power flow constraints. The main purpose of both of them is to convexify the optimization model. Specifically, the former utilizes linear equalities to approximate the power flow equations, thus transforming the nonlinear power flow constraints into linear equality constraints. However, there are always some errors between the optimal solutions obtained by the convexified model and the global optimal solutions of the original model. In addition, the optimality will be affected by the accuracy of linearization. The latter transforms the nonlinear equality constraints into convex inequality constraints, so the exact solution can be obtained with the convexified model, but the objective function and feasible domain of the optimization are required to satisfy certain conditions. Otherwise, the solution may turn out to have a substantial error.

Even though the linear power flow (LPF) is unable to guarantee exact solutions, most system operators and power industries still prefer linearization methods instead of convex relaxation methods [4]. One of the most widely used LPF model is the direct current power flow (DCPF). Although the DCPF model has a good accuracy for transmission network analysis, it is not directly applicable to distribution networks [5]. This model is not suitable for applications where reactive branch flows, voltage magnitudes, or network losses are of primary concern, such as VAR optimization, automatic voltage control (AVC) [6], and network reconfiguration. Besides, the inaccuracy of LPF may increase security risk and operational cost for system operations. Therefore, it is imperative to develop a more accurate LPF model.

### B. Literature review

Considering the characteristics of distribution systems, many LPF models and convex relaxation methods are developed:

On applicability, there are warm-start models [7-10] and cold-start models [3, 10-13]. The former kind of methods linearizes the ACPF model around the operating points, thus requiring pre-determined initial points. Cold-start models, on the contrary, do not assume initial operation points. Given a proper initial point, warm-start models can generate satisfactory power flow results. However, as the initial points are usually provided by cold-start models, the accuracy of cold-start models plays a decisive role in the results of warm-start models. Besides, in situations such as network reconfiguration where no reliable initial point is available, a cold-start model is necessary.

On the selection of state variables, there are bus injection models and branch flow models. The former is a standard model for power flow analysis and optimization. It focuses primarily on nodal variables rather than branch flows [14, 15]. Most existing LPF models fall into this category, such as [3, 9-13]. For linear branch-flow models, it uses branch flows as state variables, so the accuracy of power flow results is related to the linearization of branch flow equations. A widely-used one is the Simplified DistFlow (SD) model [16]. With directly neglecting network loss terms in the DistFlow [16], SD has cold-start and linear properties. A warm-start three-phase LBF model with modeling of on-load-tap-changer of transformers is developed in [7]. This model can produce an accurate power flow result if a proper initial point is given. Another warm-start LBF model is proposed in [8] for solving the coordinated charging issues for electric vehicles.

Generally, linearized power flow models are based on certain assumptions, including network parameter-related assumptions and operating state-related assumptions. In the derivation of DCPF, the reciprocal of reactance was used to approximate the susceptance [17], where the low *r/x* ratio network is required. To improve LPF's accuracy for different kinds of networks, a constant branch *r/x* ratio assumption for an entire system was made [11]. In [3] and [10], the voltage magnitude was approximated as 1.0 p.u. to derive the nodal power equations and branch equations. However, for practical distribution systems under variable operating conditions, these assumptions may not hold, thus limiting the applicability of these methods.

There are also studies based on the optimal power flow (OPF) model with convex relaxation techniques employed to ensure tractability. A semidefinite programming (SDP) relaxation approach based on the bus injection model is proposed in [18], which is effective when rank-1 solution is available. Such method has been widely applied to state estimation problems [19, 20]. Paper [14, 15] set forth a second-order cone program (SOCP) method that uses the reactive power capability of solar plant inverters to regulate the voltage and minimize the network loss. Paper [1] formulated the VAR optimization and network reconfiguration problem as a mixed-integer SOCP. However, considering the high DG penetration and the control of VAR compensators, SOCP cannot guarantee a feasible solution if the power injections are within a limited range, nor the objectives violate the relaxation condition [1, 21, 22]. Therefore, the existing convex relaxation methods are supposed to be improved before its practical application.

### C. Contribution

As for VAR optimization and network reconfiguration, it is not trivial to set a good initial point in all cases. Thus cold-start models become more attractive. However, current cold-start LBF models, such as simplified DistFlow, may not guarantee the optimal solution due to significant errors. There is still a gap towards an ideal approach for optimizing the operations in active distribution networks. To bridge this gap, this paper proposes a cold-start linear branch flow model, referred to as modified DistFlow, for distribution systems. Based on the modified DistFlow, a mixed-integer quadratic programming (MIQP) is proposed for VAR optimization and network reconfiguration and compared with existing benchmark. The major contributions of the paper are three-fold:

(i) To reduce the errors caused by directly neglecting network loss terms in the linearization of branch flow equations, we take the ratios of the active and reactive power to voltage magnitude as state variables, so the branch flow linearization ignores some smaller terms instead of the network loss. Furthermore, combining with the linearization of voltage differences, the equations in modified DistFlow are fully linearized in a cold-start manner for optimizations with small errors in a wide range of system states. Compared with some other cold-start models, which require assumptions of *r/x* ratio [11, 17] or voltage magnitudes around 1.0 p.u. [3, 10], the proposed model relies on much weaker assumptions. To the best of our knowledge, this is the first work of setting P/V and Q/V as state variables to establish a cold-start LBF model for distribution networks. According to our analytical analysis and simulations, the modified DistFlow is more accurate than all the existing cold-start LBF models in the literature.

(ii) By introducing the path-branch incidence matrix, we can better apply the proposed linear branch flow model, i.e., modified DistFlow, to complicated radial distribution networks. In particular, the branch flow equations in modified DistFlow can be written in a succinct matrix form.

(iii) With the modified DistFlow, the VAR optimization and network reconfiguration problem can be formulated as a MIQP problem. Compared with the MISOCP model in [1], our method can be applied to a wider range of issues and is much more efficient to solve.

The remainder of this paper is organized as follows. Section II derives the cold-start LBF model, first for a simple two-bus system, then for general distribution networks. Section III describes the MIQP model for VAR optimization and network reconfiguration. In Section IV, the proposed LBF model and the MIQP model for VAR optimization and network reconfiguration are compared with existing benchmarks on multiple standard and modified test systems.

## II. Linear Branch Flow Model

### A. A two-terminal system

*1) Branch flow equations of a two-terminal system.*

In this section, the relationship between the sending-end power flow $P_{ij}$, $Q_{ij}$ and the receiving-end power flow $P_{ji}$, $Q_{ji}$ of two connected buses are analyzed to derive and simplify the branch flow equations. Consider a single distribution line in Fig.1.

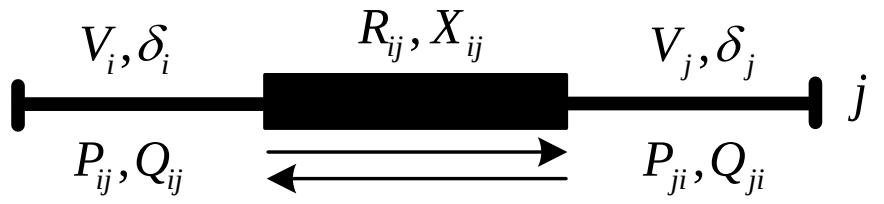

Fig. 1. A two-terminal system.

Given the sending-end power flow and receiving-end power flow, the voltage drop along the distribution line can be calculated with either bus *i* or bus *j* as the phase angle reference.

As shown in Fig. 2, for the former case, components along the horizontal and vertical directions can be calculated as:

$$\Delta V_i = \frac{R_{ij}P_{ij} + X_{ij}Q_{ij}}{V_i}, \tag{1}$$

$$\delta V_i = \frac{X_{ij}P_{ij} - R_{ij}Q_{ij}}{V_i}. \tag{2}$$

where $\mathbf{dV_i}$ is the voltage drop calculated by the sending-end power flow. $\mathbf{\Delta V_i}$ and $\boldsymbol{\delta V_i}$ are the rectangular components of $\mathbf{dV_i}$.

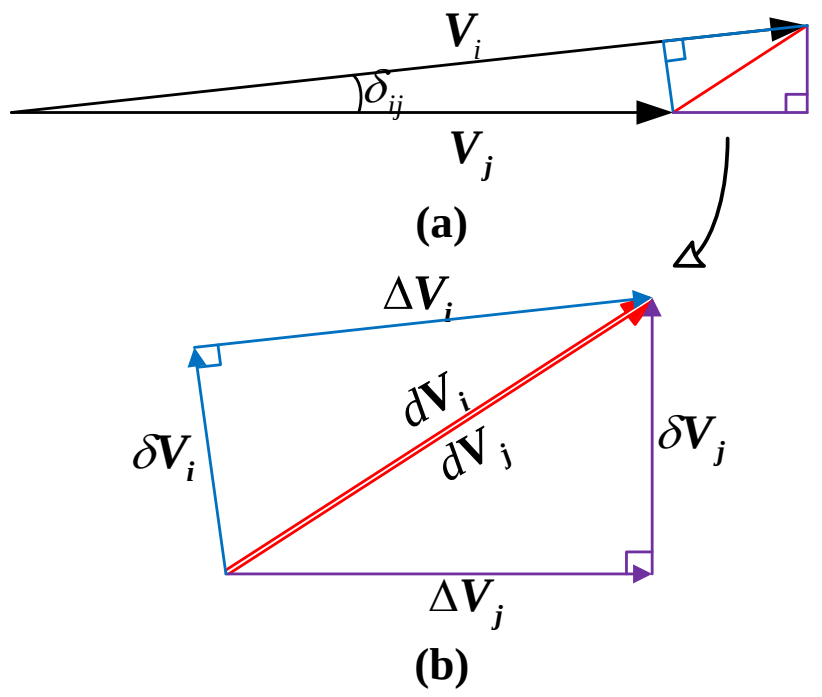

Fig. 2. Phasor diagram of the voltage drop. (Note that the figure based on the conditions $V_i > V_j$ and $\delta_{ij} > 0$ is to illustrate the following derivations (1)-(14), while the derivations are not limited to these conditions.)

In this paper, in order to obtain a linear branch flow model, the following assumption is made:

**Assumption 1.** The phase angle difference between two neighboring buses *i* and *j* is assumed to be zero. i.e., $\delta_{ij} \approx 0$.

Such an assumption is widely adopted in the literature, and verified by the simulations therein, e.g. [3, 10, 23]. Accordingly, there is:

$$\sin\delta_{ij} \approx \delta_{ij}. \tag{3}$$

The Taylor expansion of the cosine function around $\delta_{ij} = 0$ is:

$$\cos\delta_{ij} \approx 1 - \frac{1}{2}\delta_{ij}^2. \tag{4}$$

According to Fig. 2 (a), We have:

$$\Delta V_i = V_i - V_j\cos\delta_{ij} \approx V_i - V_j\left(1 - \frac{1}{2}\delta_{ij}^2\right), \tag{5}$$

$$\Delta V_j = V_i\cos\delta_{ij} - V_j \approx V_i\left(1 - \frac{1}{2}\delta_{ij}^2\right) - V_j. \tag{6}$$

By subtracting (6) from (5), we have:

$$\Delta V_i - \Delta V_j = \frac{1}{2}\left(V_i + V_j\right)\cdot\delta_{ij}^2. \tag{7}$$

Under Assumption 1, we make an approximation:

$$\Delta V_i \approx \Delta V_j. \tag{8}$$

Similarly, according to Fig. 2 (a), $\delta V_i$ and $\delta V_j$ satisfy:

$$\delta V_i = V_j \delta_{ij}, \tag{9}$$

$$\delta V_j = V_i \delta_{ij}. \tag{10}$$

By substituting (1)-(2) to (8)-(10), we obtain the relation between the power flow of the sending end and that of the receiving end:

$$\frac{P_{ij}}{V_i} = -\frac{P_{ji}}{V_j} - \frac{X_{ij}}{R_{ij}^2 + X_{ij}^2}\left(V_i - V_j\right)\delta_{ij}, \tag{11}$$

$$\frac{Q_{ij}}{V_i} = -\frac{Q_{ji}}{V_j} + \frac{R_{ij}}{R_{ij}^2 + X_{ij}^2}\left(V_i - V_j\right)\delta_{ij}. \tag{12}$$

These two equations can be simplified further. With Assumption 1, the terms containing $\delta_{ij}$ can be neglected:

$$\frac{P_{ij}}{V_i} \approx -\frac{P_{ji}}{V_j}, \tag{13}$$

$$\frac{Q_{ij}}{V_i} \approx -\frac{Q_{ji}}{V_j}. \tag{14}$$

Equations (13) and (14) show that the ratio of power flow to voltage magnitude of the sending end is approximately equal to that of the receiving end.

*Remark:*

Equations (13) and (14) form basic power flow equations of modified DistFlow. Although we assume that $\delta_{ij} \approx 0$ in Assumption 1, weaker assumptions are actually used in the derivation of (13) and (14). In Equation (8), the quadratic term of $\delta_{ij}$ is ignored. For Equations (13) and (14), the second term on the right-hand side of (11) and (12) are ignored, which are the product of ($V_i$ -$V_j$) and $\delta_{ij}$. Note that the former term is also small. Therefore, our model enjoys better accuracy than many other models with the same assumption.

For conventional methods, like simplified DistFlow [16], LPF-D [3], and [11], the branch power flows of sending end and receiving end are linearized by:

$$P_{ij} \approx -P_{ji}, \tag{15}$$

$$Q_{ij} \approx -Q_{ji}. \tag{16}$$

Then the errors $R_{ij} \cdot \left(P_{ij}^2 + Q_{ij}^2\right) / V_i^2$ and $X_{ij} \cdot \left(P_{ij}^2 + Q_{ij}^2\right) / V_i^2$ are quadratic functions of $P_{ij}$ and $Q_{ij}$. Take the two-terminal system as an example. We fix the voltage magnitude of bus *i* as 1.0 p.u., select the $P_j$ and $Q_j$ as independent variables, and plot the change of the percentage error w.r.t. $P_{ij}$ and $Q_{ij}$ calculated by conventional methods and modified DistFlow in Fig. 3.

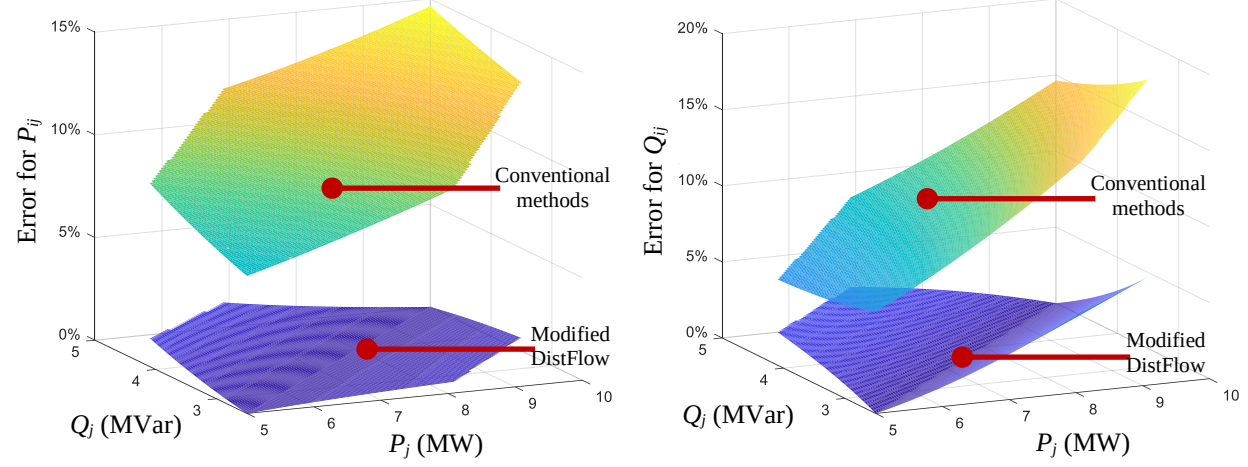

(a) (b)

Fig. 3. Visualization of branch flow errors.

From Fig.3, we can see the branch flow equations in modified DistFlow have a better accuracy than conventional methods. This is consistent with our analytical analysis: as modified DistFlow does not directly neglect network loss terms when linearizing the branch flow equations like conventional methods, the errors in branch flow results are smaller. Note that in large systems, if a branch has downstream branches, the error of downstream branches will be accumulated to the branch flow results, which makes the error of conventional methods even larger.

*2) Voltage equations of the two-terminal system.*

To visualize the non-linear nature of ACPF equations, we fix the voltage magnitude of bus *i* in the two-terminal system as 1.0 p.u. and plot the change of $V_j$ with $P_j$ and $Q_j$ as in Fig. 4 (a).

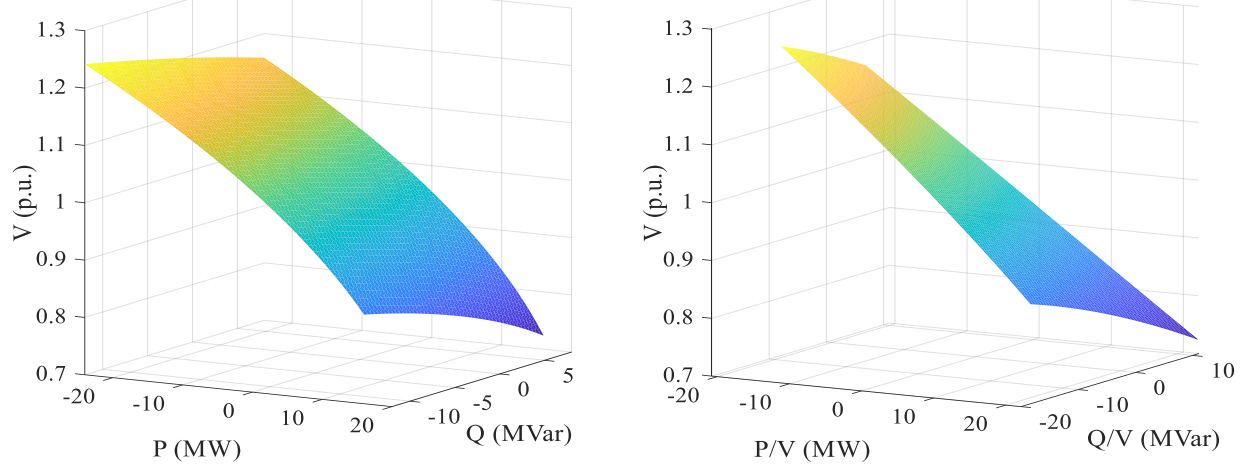

(a) (b)

Fig. 4. Visualization of ACPF in the two-terminal system.

From Fig. 4 (a) we can see there is a strong non-linearity between $V_j$ and $P_j$, $Q_j$. In contrast, if we replace $P_j$ and $Q_j$ with $P_j/V_j$ and $Q_j/V_j$, the corresponding figure in Fig. 4 (b) shows a much better linearity. Such observations inspire our idea of replacing $P_j$ and $Q_j$ with $P_j/V_j$ and $Q_j/V_j$ for the linearization of voltage equations.

The branch flow on the sending end is given by

$$P_{ij} = \frac{R_{ij}\left(V_i^2 - V_i V_j \cos\delta_{ij}\right) + X_{ij} V_i V_j \sin\delta_{ij}}{R_{ij}^2 + X_{ij}^2}, \tag{17}$$

$$Q_{ij} = \frac{X_{ij}\left(V_i^2 - V_i V_j \cos\delta_{ij}\right) - R_{ij} V_i V_j \sin\delta_{ij}}{R_{ij}^2 + X_{ij}^2}. \tag{18}$$

By combining (17) and (18), we have:

$$V_i - V_j \cos\delta_{ij} = R_{ij}\frac{P_{ij}}{V_i} + X_{ij}\frac{Q_{ij}}{V_i}, \tag{19}$$

$$V_j \sin\delta_{ij} = X_{ij}\frac{P_{ij}}{V_i} - R_{ij}\frac{Q_{ij}}{V_i}. \tag{20}$$

Under Assumption 1, (19) can be further simplified as:

$$V_i - V_j \approx R_{ij}\frac{P_{ij}}{V_i} + X_{ij}\frac{Q_{ij}}{V_i}. \tag{21}$$

Let

$$\hat{P}_{ij} = \frac{P_{ij}}{V_i},\ \hat{Q}_{ij} = \frac{Q_{ij}}{V_i}. \tag{22}$$

$$\hat{P}_i = \frac{P_i}{V_i},\ \hat{Q}_i = \frac{Q_i}{V_i}. \tag{23}$$

Then (21) becomes:

$$V_i - V_j = R_{ij}\hat{P}_{ij} + X_{ij}\hat{Q}_{ij}. \tag{24}$$

Equation (24) describes how voltage magnitude drops linearly along distribution lines. However, according to (23), $\hat{P}$ and $\hat{Q}$ are affine mappings of $W$ (i.e., $1/V$) as the $P$ and $Q$ are fixed. Thus, variables $V_i$ in (24) should be transformed to $W_i$ .

By Taylor expansion of $1/V$ at $V = 1$, a linear approximation of $W$ is obtained:

$$W = \frac{1}{V} \approx 1 + \Delta V = 2 - V. \tag{25}$$

By substituting (25) to (24), we have:

$$\frac{1}{V_j} - \frac{1}{V_i} = R_{ij}\hat{P}_{ij} + X_{ij}\hat{Q}_{ij}. \tag{26}$$

Equation (26) is the voltage equation of the proposed modified DistFlow.

According to Taylor expansion, the error for the linearization of the left-hand side of (26) is:

$$\theta(V_i, V_j) = 100 \cdot \left\| \left( \frac{1}{V_i} - (2 - V_i) \right) - \left( \frac{1}{V_j} - (2 - V_j) \right) \right\|. \tag{27}$$

Therefore, such error does not directly rely on the voltage magnitude, but the voltage difference between connected buses. Generally, the voltage drop between two buses is small, and the error introduced by (25) will be small. For example, the error for $V_i = 0.8$ and $V_j = 0.81$ is about 0.54%.

*B. Generalization*

Next, we consider general distribution networks. A simple example of a distribution feeder where bus 1 has two child-nodes is illustrated in Fig. 5.

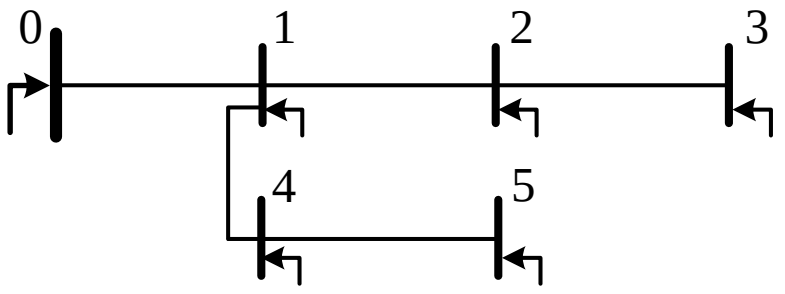

Fig. 5. A radial distribution network.

According to KCL, the sum of the injected power of a node is zero. Take bus 1 as an example. There are:

$$\frac{P_{10}}{V_1} + \frac{P_{12}}{V_1} + \frac{P_{14}}{V_1} = \frac{P_1}{V_1}, \tag{28}$$

$$\frac{Q_{10}}{V_1} + \frac{Q_{12}}{V_1} + \frac{Q_{14}}{V_1} = \frac{Q_1}{V_1}. \tag{29}$$

By substituting (13) and (14) into (28) and (29), we have

$$\frac{P_{01}}{V_0} = \frac{P_{12}}{V_1} + \frac{P_{14}}{V_1} - \frac{P_1}{V_1}, \tag{30}$$

$$\frac{Q_{01}}{V_0} = \frac{Q_{12}}{V_1} + \frac{Q_{14}}{V_1} - \frac{Q_1}{V_1}, \tag{31}$$

where (30) and (31) are the branch flow equations of the modified DistFlow.

Let $W_i$ denote $1/V_i$, and rewrite (23), (30) and (31) in the generalized form. The modified DistFlow becomes:

$$\hat{P}_{hi} = \sum_{j \in \mathrm{N}_c(i)} \hat{P}_{ij} - \hat{P}_i, \ \forall i \in \Phi_B, \tag{32}$$

$$\hat{Q}_{hi} = \sum_{j \in \mathrm{N}_c(i)} \hat{Q}_{ij} - \hat{Q}_i, \ \forall i \in \Phi_B, \tag{33}$$

$$W_j - W_i = R_{ij}\hat{P}_{ij} + X_{ij}\hat{Q}_{ij}, \ \forall ij \in \Phi_L, \tag{34}$$

$$\hat{P}_i = P_i W_i, \ \forall i \in \Phi_B, \tag{35}$$

$$\hat{Q}_i = Q_i W_i, \ \forall i \in \Phi_B, \tag{36}$$

Equations (32)-(36) are the *modified DistFlow* equations[1].

After the model is solved, the branch flow $P_{ij}$ and $Q_{ij}$ can be calculated by:

$$P_{ij} = \frac{\hat{P}_{ij}}{W_i}, \ Q_{ij} = \frac{\hat{Q}_{ij}}{W_i}. \tag{37}$$

Since the network is radial, the solution for the modified DistFlow equations can be obtained easily; for a radial network of the type shown in Fig. 5, the solutions of branch equations (32)-(33) are the following form:

$$\hat{P}_{ij} = -\sum_{k \in \Omega_{ij}} \hat{P}_k, \tag{38}$$

$$\hat{Q}_{ij} = -\sum_{k \in \Omega_{ij}} \hat{Q}_k. \tag{39}$$

Equations (32)-(33) and (38)-(39) are equivalent in many places. However, consider an optimization problem, if the network topology changes, only (32) and (33) can be employed.

To apply the proposed modified DistFlow to more general cases, the path-branch incidence matrix [24, 25] is employed. Accordingly, the modified DistFlow equations for general radial distribution networks can be written in a matrix form.

In general, the topology of a distribution network is a tree starting from a root node with multiple tributaries. The path of a specific node refers to a set of branches connecting that node to the root node. The path-branch incidence matrix describes the incidence relations between branches and paths of all nodes except the root. If the branch $l_{ij}$ belongs to the path $k$, the entry ($l_{ij}$, $k$) is 1, otherwise the entry ($l_{ij}$, $k$) is 0. Namely, its entry ($l_{ij}$, $k$) is defined as:

$$T_{l_{ij},k} = \begin{cases} 1 & l_{ij} \in \Psi_k \\ 0 & l_{ij} \notin \Psi_k \end{cases}. \tag{40}$$

Taking the simple network in Fig. 5 as an example, setting node 0 as the root, its path-branch incidence matrix is

$$T = \begin{matrix} & \begin{matrix} 1 & 2 & 3 & 4 & 5 \end{matrix} \\ \begin{matrix} l_{01} \\ l_{12} \\ l_{23} \\ l_{14} \\ l_{45} \end{matrix} & \begin{bmatrix} 1 & 1 & 1 & 1 & 1 \\ 0 & 1 & 1 & 0 & 0 \\ 0 & 0 & 1 & 0 & 0 \\ 0 & 0 & 0 & 1 & 1 \\ 0 & 0 & 0 & 0 & 1 \end{bmatrix} \end{matrix}. \tag{41}$$

For instance, elements in column 2 indicate that branch $l_{01}$ and $l_{12}$ constitute the path from node 2 to root node.

By introducing the path-branch incidence matrix, the generalized modified DistFlow equations (32) and (33) become:

$$\hat{\boldsymbol{P}}_{Br} = -\boldsymbol{T}\boldsymbol{P}_N \boldsymbol{W}_R, \tag{42}$$

$$\hat{\boldsymbol{Q}}_{Br} = -\boldsymbol{T}\boldsymbol{Q}_N \boldsymbol{W}_R. \tag{43}$$

Then, according to (32)-(36) and path-branch incidence

[1]This paper mainly focuses on three-phase balanced systems. For imbalance systems, assuming the number of nodes are the same in the three phases, the modified DistFlow can be extended to the three-phase model by (i) replacing the single-phase variables $\hat{P}_{ij}$, $\hat{Q}_{ij}$, $\hat{P}_i$, $\hat{Q}_i$, $W_i$, $P_i$, and $Q_i$ with the corresponding three-phase variables; (ii) replacing the impedance $R_{ij}$ and $X_{ij}$ with 3×3 impedance matrices, which contain the mutual impedance; (iii) changing the products of $P_i$, $Q_i$ and $W_i$ in (35) and (36) by element-wise multiplications.

matrix, the matrix form of voltage equations can be obtained:

$$\boldsymbol{W}_R = (\boldsymbol{2} - \boldsymbol{V}_0) - (\boldsymbol{T}^T \boldsymbol{R}_N \boldsymbol{T} \boldsymbol{P}_N + \boldsymbol{T}^T \boldsymbol{X}_N \boldsymbol{T} \boldsymbol{Q}_N) \boldsymbol{W}_R. \quad (44)$$

According to the approximation of $V$, $\boldsymbol{V}_R = \boldsymbol{2} - \boldsymbol{W}_R$, the voltage for each bus can be obtained by:

$$\boldsymbol{V}_R = \boldsymbol{2} - (\boldsymbol{I} + \boldsymbol{T}^T \boldsymbol{R}_N \boldsymbol{T} \boldsymbol{P}_N + \boldsymbol{T}^T \boldsymbol{X}_N \boldsymbol{T} \boldsymbol{Q}_N)^{-1} (\boldsymbol{2} - \boldsymbol{V}_0). \quad (45)$$

Equation (45) is the closed-form relation between nodal voltage magnitudes and power injections for general distribution networks.

After $\boldsymbol{V}_R$ and $\boldsymbol{W}_R$ are solved, according to (37), the branch flow $\boldsymbol{P}_{Br}$ and $\boldsymbol{Q}_{Br}$ can also be obtained by explicit forms:

$$\boldsymbol{P}_{Br} = -diag(\boldsymbol{2} - \boldsymbol{V}_S)^{-1} \boldsymbol{T} \boldsymbol{P}_N (\boldsymbol{2} - \boldsymbol{V}_R), \quad (46)$$

$$\boldsymbol{Q}_{Br} = -diag(\boldsymbol{2} - \boldsymbol{V}_S)^{-1} \boldsymbol{T} \boldsymbol{Q}_N (\boldsymbol{2} - \boldsymbol{V}_R). \quad (47)$$

## III. Applications to VAR Optimization and Network Reconfiguration

### A. Objective functions

In this section, we apply the proposed modified DistFlow model to the VAR optimization and network reconfiguration problem. Our objective is to minimize the network loss, the operating cost, the voltage magnitude deviation, or their combinations. Thus, the objective function can be written as:

$$\min_{\substack{x_{ij}, \hat{Q}_i^C, V_i, W_i, \\ \hat{P}_i, \hat{Q}_i, \hat{P}_{ij}, \hat{Q}_{ij}.}} \alpha \sum_{ij \in \Phi_L} R_{ij} \left( \hat{P}_{ij}^2 + \hat{Q}_{ij}^2 \right) + \beta \sum_{ij \in \Phi_L} (x_{ij} - x_{ij}^0)^2 + \gamma \sum_{i \in \Phi_B} \left( V_i - 1 \right)^2, \quad (48)$$

where the first term represents the network loss components. Note that the network loss of branch $ij$ is calculated by:

$$loss_{ij} = R_{ij} \frac{P_{ij}^2 + Q_{ij}^2}{V_i^2} = R_{ij} (\hat{P}_{ij}^2 + \hat{Q}_{ij}^2). \quad (49)$$

The second term denotes the cost of switch operations, and $\beta$ should be set as the switching cost. The third term represents the voltage deviation.

Decision variables herein are the status of branches $x_{ij}$, the reactive power output of shunt capacitors $\hat{Q}_i^C$, voltage magnitudes $V_i$ and its auxiliary variables $W_i$, modified power injections $\hat{P}_i$ and $\hat{Q}_i$, and modified branch flows $\hat{P}_{ij}$ and $\hat{Q}_{ij}$. The first two terms are control variables, and other variables change accordingly.

### B. Constraints

Besides, the following constraints need to be incorporated into the VAR optimization and network reconfiguration model.

*1) Integer Variable and Constraints*

Let binary variable $x_{ij}$ denote the branch switch status (where $x_{ij} = 1$ indicates that switch $ij$ is closed, whereas $x_{ij} = 0$ suggests that the switch $ij$ is open).

$$x_{ij} \in \{0,1\}, \quad \forall ij \in \Phi_L. \quad (50)$$

When a branch $ij$ is open, i.e., $x_{ij} = 0$, its active and reactive power flow must be zero. Such limitations can be written in the following form of linear inequality constraints:

$$-x_{ij} \cdot M \le \hat{P}_{ij} \le x_{ij} \cdot M, \quad (51)$$

$$-x_{ij} \cdot M \le \hat{Q}_{ij} \le x_{ij} \cdot M. \quad (52)$$

*2) Branch Flow Constraints*

According to modified DistFlow, the branch flow constraints can be written as:

$$\hat{P}_{hi} = \sum_{j \in \mathrm{N}_c(i)} \hat{P}_{ij} - \hat{P}_i, \ \forall j \in \Phi_B, \quad (53)$$

$$\hat{Q}_{hi} = \sum_{j \in \mathrm{N}_c(i)} \hat{Q}_{ij} - \hat{Q}_i, \ \forall j \in \Phi_B. \quad (54)$$

*3) Voltage Constraints*

According to modified DistFlow, for a closed branch, the voltage drop along the distribution line is:

$$W_j - W_i = R_{ij} \hat{P}_{ij} + X_{ij} \hat{Q}_{ij}, \ \forall ij \in \Phi_L, \quad (55)$$

$$V_i = 2 - W_i, \ \forall ij \in \Phi_L. \quad (56)$$

To incorporate the case with open branches, the big-M method is introduced [26] and the voltage constraints (55) become:

$$W_j - W_i \le \left(1 - x_{ij}\right) M + R_{ij} \hat{P}_{ij} + X_{ij} \hat{Q}_{ij}, \ \forall ij \in \Phi_L, \quad (57)$$

$$W_j - W_i \ge -\left(1 - x_{ij}\right) M + R_{ij} \hat{P}_{ij} + X_{ij} \hat{Q}_{ij}, \ \forall ij \in \Phi_L. \quad (58)$$

*4) Active and Reactive Power Injection Constraints*

$$\hat{P}_i = -P_i^D W_i + \tilde{P}_i^G W, \ \forall i \in \Phi_B, \quad (59)$$

$$\hat{Q}_i = -Q_i^D W_i + \tilde{Q}_i^G W + \hat{Q}_i^C, \ \forall i \in \Phi_B. \quad (60)$$

*5) VAR Compensators Operation Constraints*

The constraints for static VAR compensators (SVC) are:

$$\underline{Q}_i^C W_i \le \hat{Q}_i^C \le \overline{Q}_i^C W_i, \ \forall i \in \Phi_C. \quad (61)$$

*6) Radiation Constraints*

According to [27], the constraints for a radial topology can be obtained by:

$$\sum_{i \ne j} x_{ij} = N_{node} - N_{sub}, \ \forall ij \in \Phi_L, \quad (62)$$

$$\sum_{l_{ji} \in \Phi_L} k_{ji} - \sum_{l_{ij} \in \Phi_L} k_{ij} = K_i, \ \forall i \in \Phi_B, \quad (63)$$

$$K_i = 1, \ \forall i \in \Phi_G \cup \Phi_C, \quad (64)$$

$$K_i = 0, \ \forall i \notin \Phi_G \cup \Phi_C \cup \Phi_{Sub}, \quad (65)$$

$$\left| k_{ij} \right| \le N_G x_{ij}, \ \forall ij \in \Phi_L. \quad (66)$$

*7) Branch Power Flow Limits*

Limits on active and reactive power flow:

$$-\overline{P}_{ij} W_i \le \hat{P}_{ij} \le \overline{P}_{ij} W_i, \ \forall ij \in \Phi_L, \quad (67)$$

$$-\overline{Q}_{ij} W_i \le \hat{Q}_{ij} \le \overline{Q}_{ij} W_i, \ \forall ij \in \Phi_L, \quad (68)$$

$$\hat{P}_{ij}^2 + \hat{Q}_{ij}^2 \le \overline{I}_{ij}^2, \ \forall ij \in \Phi_L, \quad (69)$$

where (67) and (68) are the capacity constraints for branches, and (69) are the thermal limit constraints. In practical applications, if the utilities do not prefer the reverse power follow appears in their grids, the lower bound of constraints (67) and (68) can be set to 0.

*8) Bus Voltage Limits*

$$2 - \overline{V}_i \le W_i \le 2 - \underline{V}_i, \ \forall i \in \Phi_B. \quad (70)$$

*9) Phase Angle Difference Limits*

Since the modified DistFlow is based on Assumption 1, we make the following phase angle constraints according to (20):

$$-\left(2 - W_j\right) \sin \overline{\delta}_{ij} \le X_{ij} \hat{P}_{ij} - R_{ij} \hat{Q}_{ij} \le \left(2 - W_j\right) \sin \overline{\delta}_{ij}, \ \forall ij \in \Phi_L. \quad (71)$$

We usually set $\overline{\delta}_{ij}$ to 10 degrees.

In this paper, we consider constraints *1)-9)*. However, other types of constraints can also be considered as long as they can be written as convex constraints. For example, the power factor constraint at the feeder root:

$$\frac{P_{01}^2}{P_{01}^2+Q_{01}^2} \ge \underline{\eta}. \tag{72}$$

Then, assuming the active power generation at the root node is always larger than 0. According to the variable definition in modified DistFlow, the power factor constraint can be transformed into a convex inequality constraint:

$$\hat{P}_{01} - \sqrt{\frac{\underline{\eta}}{1-\underline{\eta}}}\left|\hat{Q}_{01}\right| \ge 0. \tag{73}$$

Note that the objective function (48) is a convex quadratic function. Equality constraints {(53), (54), (56), (59), (60), (62)} are affine and inequality constraints {(51), (52), (57), (58), (61), (67)-(70)} are convex. Thus the VAR optimization and network reconfiguration problem becomes a MIQP with the proposed modified DistFlow model.

### C. *Method to improve computational efficiency*

The mixed-integer programming is usually solved by branch and cut algorithms, which suffers from a high computation burden in dealing with integer variables. The computation time for network reconfiguration will be reduced if the status of some branches can be predetermined.

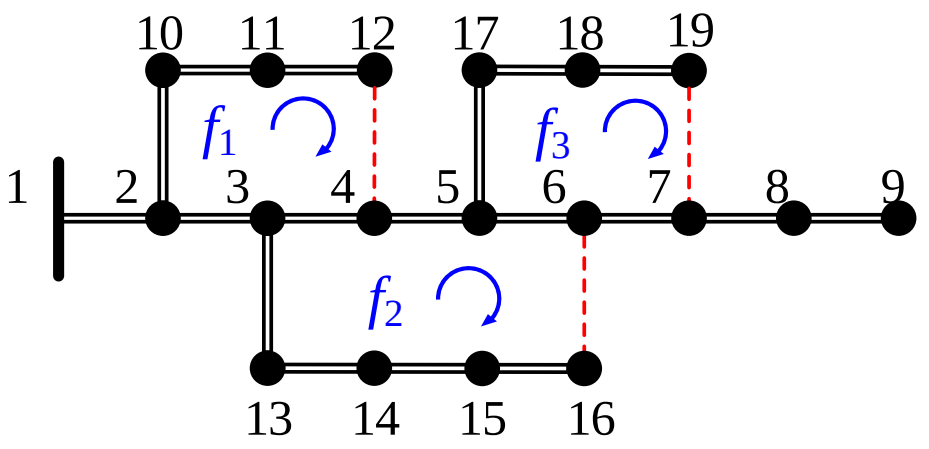

Fig. 6. A simple distribution system.

A simple distribution network is shown in Fig. 6 to introduce the basic concepts of the method. The normally closed branches are shown as double solid lines, and the normally opened branches are displayed as red dashed lines.

Define two loops $f_1$ and $f_2$ are *overlapping* if $B_{f1} \cap B_{f2} \neq \emptyset$. If the two loops $f_1$ and $f_2$ are not overlapped with other loops, {$f_1$, $f_2$} is the overlapping loop set. For the $k$th overlapping loop set, there are $L_k$ branches in $\Theta_k$ opened otherwise there will be a loop in the network, meanwhile, the isolated island will appear at other places.

$$N_k - \sum x_{ij} = L_k, \ \forall ij \in \Theta_k. \tag{74}$$

Constraints (74) are necessary conditions for radial configuration, which could save time to judge whether a solution is feasible. This method is strict in mathematics and will not affect the optimality of the results.

## IV. Numerical Tests

We tested the proposed modified DistFlow model and the VAR optimization and network reconfiguration technique, respectively. In all tests, the voltage magnitude of the power supply point (PSP) was set as 1.05 p.u. All simulations were tested in MATLAB on a laptop with an Intel Core i7-5600U 2.60GHz CPU and 8GB of RAM.

### A. *On the linear branch flow model*

There are two categories of cold-start linear power flow model, i.e. bus injection model and branch flow model. In each category, we pick one presentative work as the benchmark in our comparison. For linear bus injection model, we pick LPF-D [3] as one of our benchmarks. For many other linear models in this category, their linearization of branch power flows is the same, which may lead to the same error in branch flow results. For linear branch flow model, we choose simplified DistFlow [16] as the other benchmark, which is one of the most widely used models in this category.

Therefore, the proposed modified DistFlow model was compared with multiple benchmarks, including LPF-D [3], simplified DistFlow [16], as well as the ACPF model calculated by MATPOWER 7.1 [28]. We first chose the 33-bus system as our testbed. The results of voltage magnitudes, branch active power flows, and branch reactive power flows calculated by ACPF as well as the errors of modified DistFlow (MD), LPF-D, and SD were illustrated in Fig. 7-9, respectively.

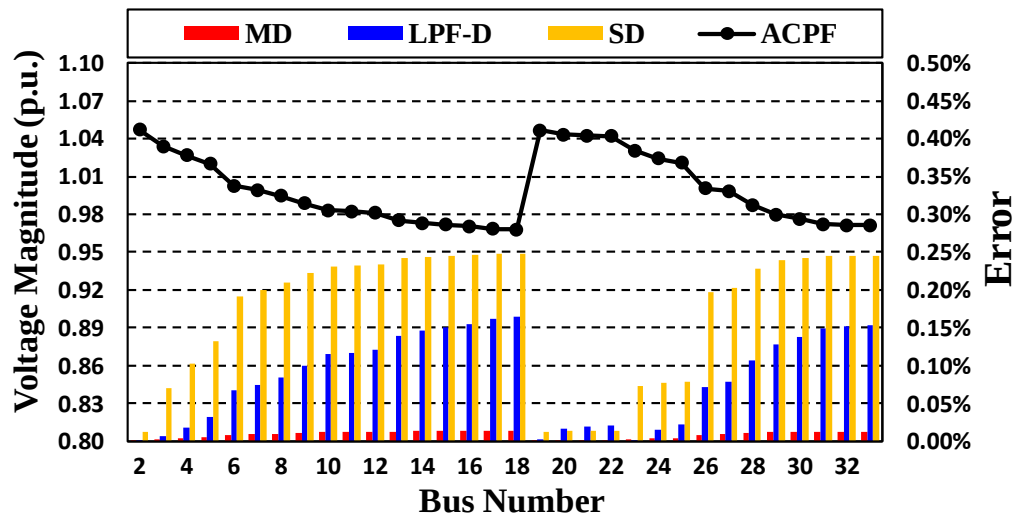

Fig. 7. Results of bus voltage magnitudes.

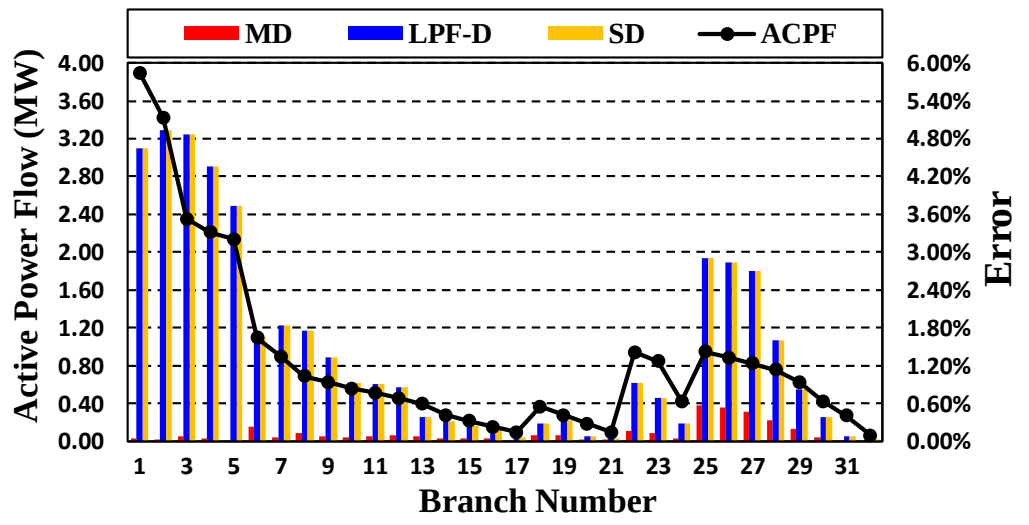

Fig. 8. Results of branch active power flows.

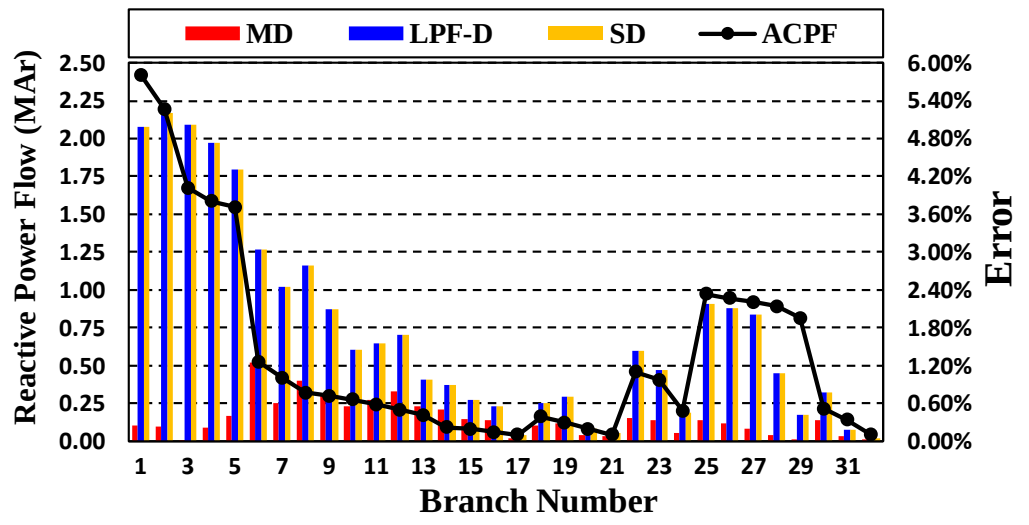

Fig. 9. Results of branch reactive power flows.

Fig. 7 shows that the voltage magnitude results of MD are closest to ACPF. The average error and the largest error of MD were 0.008% and 0.014%, respectively. The average error and the largest error of LPF-D were 0.085% and 0.164%, respectively. The average error and the largest error of SD were 0.170% and 0.247% respectively. We also observed that errors grew larger when the voltage magnitude became lower.

However, even the largest errors of MD were within a negligible range and were much smaller than LPF-D and SD.

As shown in Figs. 8 and 9, the results of active and reactive branch flow from MD were also closest to ACPF. The largest errors were 0.559% at branch 25 and 1.236% at branch 6 for active and reactive power flow, respectively. The average errors were 0.118% and 0.351% for active and reactive power flow, respectively. Compared to LPF-D and SD, the branch flow results of MD were much more accurate. Meanwhile, from Fig. 8-9, we observed that errors on active and reactive power flows for LPF-D and SD were the same, this is because they ignoring the network loss when linearizing the branch flow equations. (Please refer to (16)~(27) in [3] and (5.i, 5.ii) in [16]).

Subsequently, the proposed modified DistFlow model was tested and compared based on the 141-bus system. Its mean and largest errors for voltage magnitudes, branch active power flows, and branch reactive power flows were recorded in TABLE I, along with the results of LPF-D and SD. From TABLE I, we can see that errors of MD were much smaller than those of LPF-D and SD.

TABLE I
POWER FLOW RESULTS COMPARISON ON 141-BUS SYSTEM

| Error | Method | Bus Voltages | Branch Flows | |
|---|---|---|---|---|
| | | $V_i$ | $P_{ij}$ | $Q_{ij}$ |
| Average Error | MD | 0.002% | 0.024% | 0.044% |
| | LPF-D | 0.024% | 0.334% | 0.394% |
| | SD | 0.129% | 0.334% | 0.394% |
| Largest Error | MD | 0.003% | 0.471% | 0.407% |
| | LPF-D | 0.064% | 4.522% | 5.350% |
| | SD | 0.178% | 4.522% | 5.350% |

The increasing electric vehicle charging demand potentially leads to overload [29]. To check the performance of the proposed modified DistFlow model heavier load conditions, we employed the 33-bus system and 141-bus system, and gradually increased the load power therein. Individually, loads of 33-bus systems were scaled up to 210%~250% of the baseload, and loads of 141-bus systems were scaled up to 260%~300% of the baseload. These values would gradually make the lowest voltage magnitude in the system close to 0.8 p.u. The mean and largest errors in voltage magnitudes, branch active power flows, and branch reactive power flows for all linear power flow models were compared in TABLE II.

In Table II, the lowest voltage decreased from 0.86 p.u. to as low as 0.81 p.u. From the table, it can be observed that the voltage error of MD was less affected by low voltage. When the voltage magnitude was around 0.8 p.u., the average and the largest errors of voltage magnitude were about 0.5% and 1%, respectively, and the errors of branch power flows were still acceptable. Whereas for LPF-D and SD, the errors of voltage magnitudes and branch power flows were several times as much as modified DistFlow, which verified the better performance of modified DistFlow under low voltage conditions.

## B. *VAR optimization and network reconfiguration*

Various cases considering DGs and SVCs were used to compare the performance of the MIQP model and MISOCP model [1] for VAR optimization and network reconfiguration problems, and the algorithms were solved by an embedded IBM CPLEX 12.8 solver with the YALMIP interface in MATLAB.

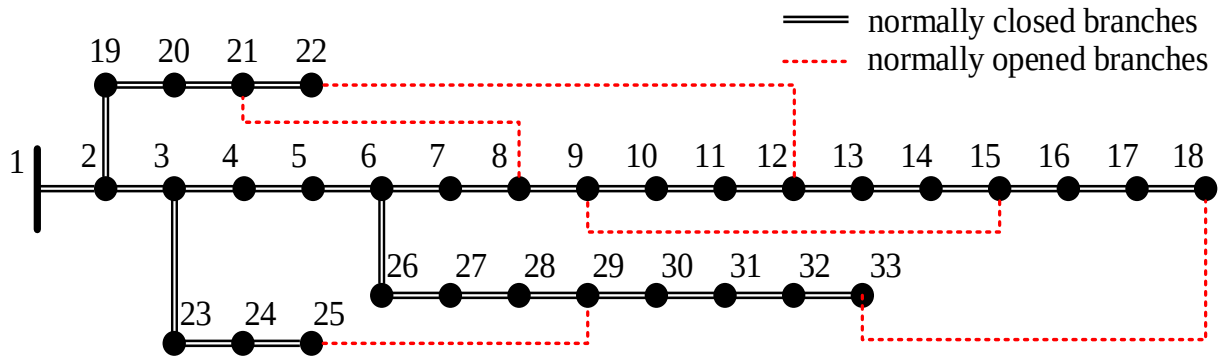

Fig. 10. The 33-bus system.

We first chose the 33-bus system as the testbed. In the VAR optimization and network reconfiguration, three objectives and three cases were tested to compare the MIQP model and the MISOCP model [1]. Note that the proposed method to improve computational efficiency hasn't been used to the modified DistFlow and benchmarks in the 33-bus system case, so that we can compare our method with the SOCP-based method fairly.

*Objective I*: Minimizing the active power losses.

*Objective II*: Minimizing the cost of network reconfiguration. The energy cost $\alpha$ was set as 30$/MWh, and the switching cost $\beta$ was estimated as 0.2$.

*Objective III*: Minimizing the voltage deviation. The coefficient $\alpha$ and $\beta$ were set as 0, and the $\gamma$ was set as 100. To highlight the effect of network reconfiguration, loads of the whole system were scaled up to 150% of the baseload.

To diverse the testbed, we set three cases:

*Case I*: There was no DG and SVC in the system.

TABLE II
COMPARISONS AMONG MODIFIED DISTFLOW, LPF-D AND SIMPLIFIED DISTFLOW UNDER HEAVY LOAD SYSTEM

| System | Load Level | Lowest $V$ (p.u.) | Modified DistFlow | | | | | | LPF-D | | | | | | Simplified DistFlow | | | | | |
|---|---|---|---|---|---|---|---|---|---|---|---|---|---|---|---|---|---|---|---|---|
| | | | Error $V$ (%) | | Error $P_{ij}$ (%) | | Error $Q_{ij}$ (%) | | Error $V$ (%) | | Error $P_{ij}$ (%) | | Error $Q_{ij}$ (%) | | Error $V$ (%) | | Error $P_{ij}$ (%) | | Error $Q_{ij}$ (%) | |
| | | | $\varepsilon_V^{avg}$ | $\varepsilon_V^{max}$ | $\varepsilon_P^{avg}$ | $\varepsilon_P^{max}$ | $\varepsilon_Q^{avg}$ | $\varepsilon_Q^{max}$ | $\varepsilon_V^{avg}$ | $\varepsilon_V^{max}$ | $\varepsilon_P^{avg}$ | $\varepsilon_P^{max}$ | $\varepsilon_Q^{avg}$ | $\varepsilon_Q^{max}$ | $\varepsilon_V^{avg}$ | $\varepsilon_V^{max}$ | $\varepsilon_P^{avg}$ | $\varepsilon_P^{max}$ | $\varepsilon_Q^{avg}$ | $\varepsilon_Q^{max}$ |
| 33-Bus System | 210% | 0.860 | 0.213 | 0.397 | 0.615 | 2.359 | 1.170 | 3.766 | 1.324 | 2.455 | 3.581 | 11.601 | 4.253 | 12.187 | 1.088 | 1.681 | 3.581 | 11.601 | 4.253 | 12.187 |
| | 220% | 0.849 | 0.266 | 0.496 | 0.709 | 2.623 | 1.305 | 4.093 | 1.541 | 2.861 | 3.814 | 12.325 | 4.529 | 12.921 | 1.244 | 1.934 | 3.814 | 12.325 | 4.529 | 12.921 |
| | 230% | 0.837 | 0.330 | 0.617 | 0.814 | 2.909 | 1.453 | 4.443 | 1.783 | 3.316 | 4.057 | 13.073 | 4.816 | 13.677 | 1.419 | 2.220 | 4.057 | 13.073 | 4.816 | 13.677 |
| | 240% | 0.825 | 0.406 | 0.762 | 0.930 | 3.221 | 1.614 | 4.817 | 2.054 | 3.827 | 4.311 | 13.848 | 5.116 | 14.458 | 1.615 | 2.544 | 4.311 | 13.848 | 5.116 | 14.458 |
| | 250% | 0.813 | 0.497 | 0.938 | 1.060 | 3.562 | 1.790 | 5.218 | 2.357 | 4.401 | 4.576 | 14.654 | 5.429 | 15.266 | 1.835 | 2.912 | 4.576 | 14.654 | 5.429 | 15.266 |
| 141-Bus System | 260% | 0.850 | 0.237 | 0.466 | 0.133 | 1.657 | 0.315 | 3.173 | 1.611 | 2.934 | 1.046 | 13.359 | 1.217 | 15.500 | 1.350 | 2.042 | 1.046 | 13.359 | 1.217 | 15.500 |
| | 270% | 0.840 | 0.287 | 0.565 | 0.154 | 1.917 | 0.346 | 3.509 | 1.823 | 3.325 | 1.102 | 14.013 | 1.281 | 16.236 | 1.506 | 2.295 | 1.102 | 14.013 | 1.281 | 16.236 |
| | 280% | 0.830 | 0.346 | 0.682 | 0.176 | 2.203 | 0.379 | 3.873 | 2.056 | 3.756 | 1.160 | 14.684 | 1.347 | 16.989 | 1.678 | 2.577 | 1.160 | 14.684 | 1.347 | 16.989 |
| | 290% | 0.820 | 0.415 | 0.820 | 0.202 | 2.517 | 0.414 | 4.268 | 2.310 | 4.232 | 1.220 | 15.374 | 1.416 | 17.761 | 1.866 | 2.890 | 1.220 | 15.374 | 1.416 | 17.761 |
| | 300% | 0.809 | 0.495 | 0.982 | 0.229 | 2.862 | 0.452 | 4.695 | 2.590 | 4.758 | 1.282 | 16.084 | 1.486 | 18.552 | 2.074 | 3.239 | 1.282 | 16.084 | 1.486 | 18.552 |

TABLE III
RESULTS OF 33-BUS SYSTEM

| NO. | Objective function | Load Level | Case | MIQP Obj.[1] | MIQP Loss(kW) | MIQP $V_{avg}$(p.u.) | MIQP Opened Branches | MIQP Time(s) | MISOCP Obj.[1] | MISOCP Loss(kW) | MISOCP $V_{avg}$(p.u.) | MISOCP Opened Branches | MISOCP Time(s) |
|---|---|---|---|---|---|---|---|---|---|---|---|---|---|
| 1 | Obj. 1 | 100% | I | 125.43 | 125.43 | 1.017 | 7-8, 9-10, 14-15, 32-33, 25-29 | 3.64 | 125.43 | 125.43 | 1.017 | 7-8, 9-10, 14-15, 32-33, 25-29 | 7.66 |
| 2 | | | II | 81.93 | 81.93 | 1.024 | 6-7, 8-9, 14-15, 12-22, 25-29 | 3.63 | 81.93 | 81.93 | 1.024 | 6-7, 8-9, 14-15, 12-22, 25-29 | 6.10 |
| 3 | | | III | 53.07 | 53.07 | 1.032 | 7-8, 10-11, 14-15, 9-15, 25-29 | 3.20 | 53.07 | 53.07 | 1.032 | 7-8, 10-11, 14-15, 9-15, 25-29 | 8.18 |
| 4 | Obj. 2 | 100% | I | 4.53 | 137.79 | 1.014 | 8-9, 21-8, 9-15, 18-33, 25-29 | 0.95 | 4.53 | 137.79 | 1.014 | 8-9, 21-8, 9-15, 18-33, 25-29 | 2.28 |
| 5 | $C^P = 30$ | | II | 3.04 | 101.41 | 1.019 | 21-8, 9-15,12-22, 18-33,25-29 | 0.50 | 3.04 | 101.41 | 1.019 | 21-8, 9-15,12-22, 18-33,25-29 | 2.01 |
| 6 | $C^{cb} = 0.2$ | | III | 2.15 | 71.86 | 1.024 | 21-8, 9-15,12-22, 18-33,25-29 | 0.28 | 2.15 | 71.86 | 1.024 | 21-8, 9-15,12-22, 18-33,25-29 | 0.89 |
| 7 | Obj. 3 | 150% | I | 1.86 | 295.51 | 1.000 | 7-8, 9-10, 14-15, 32-33, 25-29 | 16.69 | NOT APPLICABLE (infeasible power flow solution) | | | | - |
| 8 | $\alpha = 0$ | | II | 1.46 | 260.98 | 1.003 | 4-5, 10-11,14-15, 28-29,32-33 | 16.39 | NOT APPLICABLE (infeasible power flow solution) | | | | - |
| 9 | $\beta = 100$ | | III | 1.11 | 251.23 | 0.998 | 4-5, 8-9, 14-15, 27-28, 32-33 | 6.27 | NOT APPLICABLE (infeasible power flow solution) | | | | - |

[1] the values of objective function and loss were calculated by ACPF results based on the branches' status that had been obtained by the mixed-integer problem.

*Case II*: There was only one DG installed at Bus 10 with an output of 0.8 MW and 0.5 MVar.

*Case III*: There were two DGs at Bus 16 and 30 with an output of 0.5 MW and 0.25 MVar, and an SVC at Bus 22 within an output range of [-0.5, 0.5] MVar.

Combined with the objective functions and cases, there were totally nine scenarios. In these scenarios, we compared the MIQP proposed in this paper with the MISOCP proposed in [1] from the aspects of optimal value, optimal solution, and solution time. The comparison results were shown in Table III.

In scenarios 1-6, according to [14], minimizing the network losses as the objective function satisfies the relaxation condition of SOCP. Therefore, the solutions obtained by MISOCP, which are usually optimal, can be regarded as benchmarks. From Table III, we can see that the optimal value of MIQP is the same as MISOCP, and the branch status in the optimal solution of MIQP is consistent with that of MISOCP, which verifies the optimality and accuracy of MIQP. Meanwhile, it can be observed that the time costs of MIQP were much lower than MISOCP in all scenarios.

In scenarios 7-9, because the objective function of minimizing the voltage deviation does not satisfy the relaxation condition of SOCP, MISOCP is not applicable to this problem. While MIQP still solved these problems successfully.

From these results, we conclude that i) when the MISOCP model is applicable, the proposed MIQP model obtained the same results with much lower time costs, and ii) when the MISOCP model is not applicable, the proposed MIQP approach still solved the VAR optimization and network reconfiguration problem successfully.

To test the performance of the proposed MIQP model and the acceleration method on large systems, a 981-bus system was adopted for testing, which combined seven 141-bus systems with different load levels and inter-area switch connections. We set up six normally open lines, which were represented by the red dashed line in the figure. Therefore, there were 986 lines in the system. To ensure the feasible solution of power flow calculation, we reduced the load in each system according to the ratio shown in Fig. 11.

The proposed MIQP model and the acceleration method were used to calculate the optimal topology and network losses. Subsequently, we used ACPF to calculate the power flow under the optimal topology to verify the accuracy of the MIQP model optimal solution. The result shows that it took 233.4 seconds to obtain the optimal solution, which confirmed that the proposed model and algorithm could be applied to VAR optimization network reconfiguration in large systems. In contrast, the MISOCP model failed to solve the VAR optimization and network reconfiguration issues for this system.

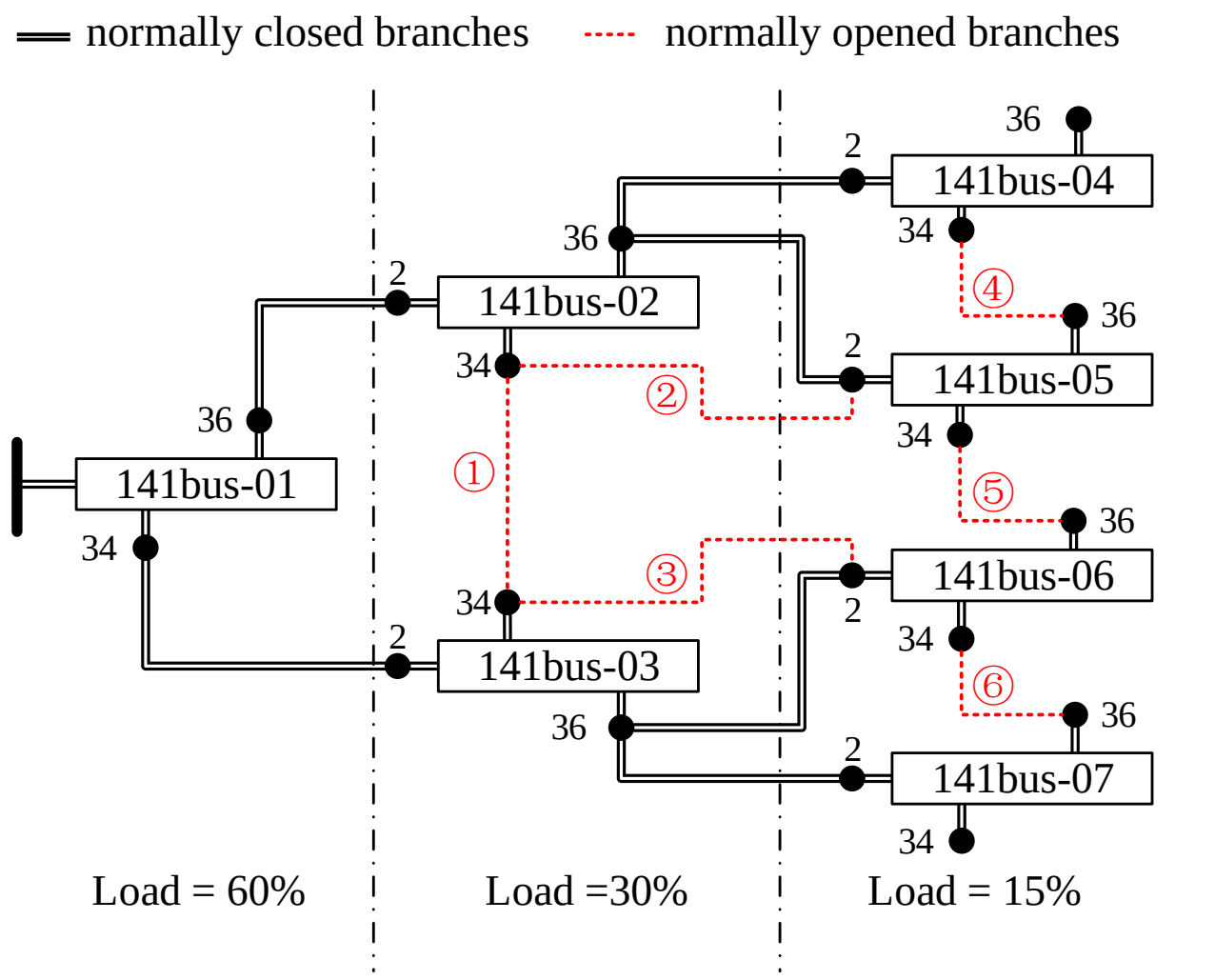

Fig. 11. A 981-bus system.

In summary, we conclude that the proposed modified DistFlow model is much more accurate than existing benchmarks, and the MIQP model based on modified DistFlow enjoys satisfactory accuracy and efficiency even for large systems.

## V. CONCLUSION

In this paper, a cold-start LBF model named modified DistFlow is proposed by replacing the active and reactive power with their ratios to voltage magnitude as the state variables. Such a LBF model is applied to the problem of VAR optimization and network reconfiguration, transforming it to a MIQP. Theoretical analysis and numerical tests both show that the proposed modified DistFlow has an outstanding accuracy, and the resulting MIQP model for VAR optimization and network reconfiguration is very efficient to solve and can be applied to a wide range of problems. We also look forward to solving more distribution system problems with the modified DistFlow model as our basic tool in the future.